# DEVELOPING AND DEPLOYING SECURITY APPLICATIONS FOR IN-VEHICLE NETWORKS

**Samuel C Hollifield**
Oak Ridge National Laboratory

**Pablo Moriano**
Oak Ridge National Laboratory

**William L Lambert**
Tennessee Technological University

**Joel Asiamah**
Oak Ridge National Laboratory

**Isaac Sikkema**
Oak Ridge National Laboratory

**Michael D Iannacone**
Oak Ridge National Laboratory

## ABSTRACT

Radiological material transportation is primarily facilitated by heavy-duty on-road vehicles. Modern vehicles have dozens of electronic control units (ECUs), which are small, embedded computers that communicate with sensors and each other for vehicle functionality. ECUs use a standardized network architecture—Controller Area Network (CAN)—which presents grave security concerns that have been exploited by researchers and hackers alike. For instance, ECUs can be impersonated by adversaries who have infiltrated an automotive CAN and disable or invoke unintended vehicle functions such as brakes, acceleration, or safety mechanisms. Further, the quality of security approaches varies wildly between manufacturers. Thus, research and development of after-market security solutions have grown remarkably in recent years. Many researchers are exploring deployable intrusion detection and prevention mechanisms using machine learning and data science techniques. However, there is a gap between developing security system algorithms and deploying prototype security appliances in-vehicle. In this paper, we, a research team at Oak Ridge National Laboratory (ORNL) working in this space, highlight challenges in the development pipeline, and provide techniques to standardize methodology and overcome technological hurdles. To bridge the computer architecture gap between standard development environments and vehicle-embedded devices, ORNL uses a CAN-based testbed with simulated ECUs to transfer the algorithms to ARM-architecture embedded computers and microcontrollers. This step typically involves solving code-dependency errors and stress-testing for in-situ deployment on streaming network data. ORNL then implements these embedded devices onto heavy-duty trucks to pilot-test the hardware and refine algorithms for a vehicle-agnostic deployment. This approach has accelerated the development of in-vehicle security technologies while also decreasing the cost of labor and increasing programming and security efficiencies over developing CAN IDS technologies directly on vehicles.

## INTRODUCTION

Radiological material is transported using on-road vehicles. Modern vehicles are reliant on electronic control units (ECUs), which are small, embedded computers that have specific functions and operate many vehicle components. These ECUs must communicate with each other to synchronize

This manuscript has been co-authored by UT-Battelle, LLC, under contract DE-AC05-00OR22725 with the US Department of Energy (DOE). The US government retains and the publisher, by accepting the article for publication, acknowledges that the US government retains a nonexclusive, paid-up, irrevocable, worldwide license to publish or reproduce the published form of this manuscript, or allow others to do so, for US government purposes. DOE will provide public access to these results of federally sponsored research in accordance with the DOE Public Access Plan (http://energy.gov/downloads/doe-public-access-plan).

vehicle systems and provide critical sensor readings or inputs; thus, a shared network architecture is used. The network type commonly used in vehicles is Controller Area Network (CAN) [1]. CAN is a real-time, embedded network that is easy to implement and allows error handling and correction. However, recent research has demonstrated that CAN harbors inherent vulnerabilities that severely impact the safety and security of automobiles. E.g., in 2015, security researchers compromised an internet-connected infotainment unit in a Chrysler vehicle, resulting in the recall of 1.4 million automobiles [2]. Attacks that leverage CAN-based vulnerabilities have also been launched recently by thieves, using modified CAN hardware to steal Toyota and Lexus vehicles [3]. Further, many attacks documented within the research community have exploited CANs error handling, arbitration, and applications [4][5].

Although attention to transportation cybersecurity has grown in recent years, vehicle manufacturers need to be more consistent in their application of defensive tools. As a result, third-party researchers and industry stakeholders began exploring after-market cybersecurity solutions. This typically involves algorithmic development to establish a detection scheme, transferring the detection capabilities to embedded hardware, then deploying a prototype defensive tool for real-time vehicle monitoring. We have utilized this approach at Oak Ridge National Laboratory (ORNL), where our research involves rapidly prototyping intrusion detection systems for CAN-based automotive networks.

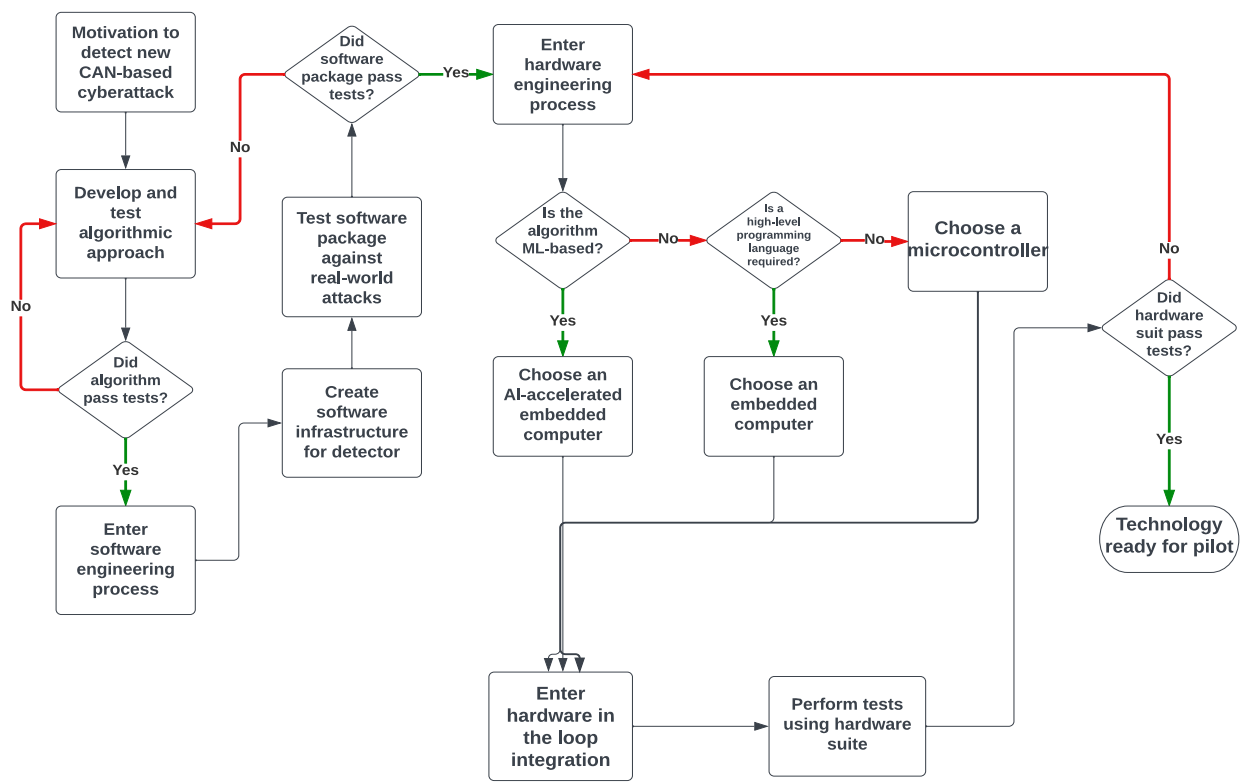

**Figure 1. In-vehicle network security IDS pipeline depicted.**

Our research and development pipeline, depicted in Figure 1 above, uses open-source, freely available tools to create custom solutions that circumvent prohibitively costly, manufacturer-targeted tools to accomplish our goal. Throughout our research, we have encountered numerous hurdles that require solutions that, when known, streamline the process. First, embedded hardware has a different computational ability than the powerful computers used for machine learning and data science. Choosing the appropriate hardware for the project is critical in ensuring the software can function correctly. Second, testbench development, including hardware-in-the-loop integration, is crucial to reduce development times. Finally, prototype deployment includes considerations for environmental protection, power and network access, and secure mounting.

**INTRODUCTION**

CAN 2.0, the broadcast communication protocol currently used in automobiles, was published in 1991 to define CAN communications' physical and logical characteristics. CAN communication is facilitated by both a CAN controller and a CAN transceiver. These are pieces of hardware that enable host devices to connect and process CAN messages. The physical transmission of CAN occurs through a pair of twisted wires connected to every CAN transceiver on the network. Transmitted logical bits are arranged into CAN frames per the CAN 2.0 specification, as Figure 2 below depicts. Although CAN frames have many components, only two are used for vehicle function:

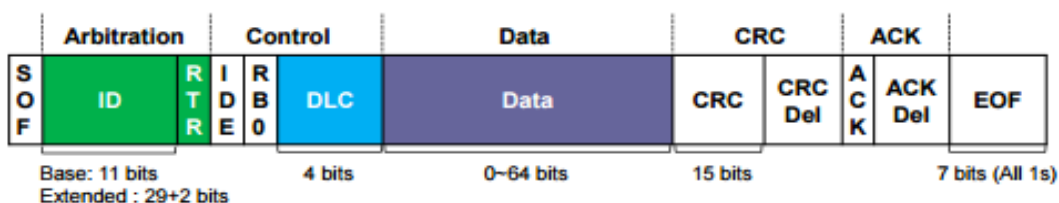

**Figure 2. Basic CAN frame depicted. Photo credit, Cho & Shin [4].**

- *Arbitration ID (AID):* Either 11 or 29 bits in length, AID is used to identify the packet contents and determine the message's priority compared to others transmitting simultaneously.
- *Data*: Up to 64 data bits can be transmitted in one message. The encoding of data fields is generally held secret by automobile manufacturers and varies per make, model, year, and trim.

Understanding a CAN data frame's meaning burdens independent cybersecurity research. A single CAN payload may include multiple signals of different data types, sizes, and scaling. These signals often do not follow a delimiting pattern at the byte level—meaning it is nontrivial to decode complex signals that exceed byte boundaries. An example of a CAN payload's signal extraction is represented in Figure 3 below. This 8 x 8 array represents the 64 bits within a CAN message payload, indicating each signal's representative bits with a unique color. Unused bits are shown in white.

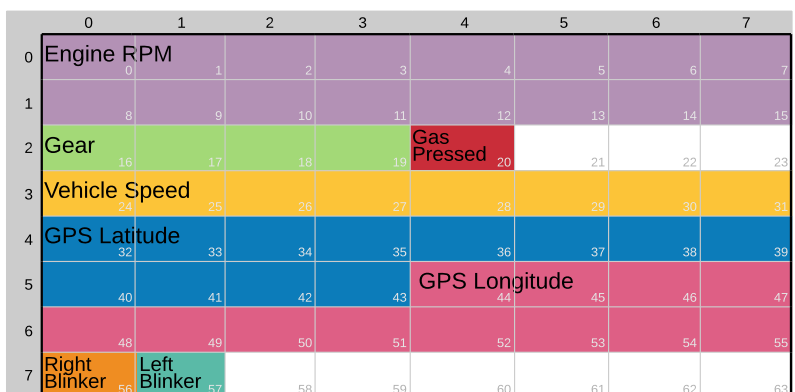

**Figure 3. CAN signal plot, 8 x 8 array of signals encoded in an AID's data frame depicted. Photo credit, Verma et al. [6].**

Although CAN provides incredible resilience for embedded networking, it is bereft of security features critical for secure communications. All connected devices can read messages regardless of the intended recipient. There is no inherent authentication to verify the integrity of a message's source, meaning spoofing and message injection will often cause a vehicle to behave unexpectedly or unsafely. This is particularly concerning for heavy-duty vehicles, which contain a much larger kinetic energy footprint and are used for various cargo, e.g., transporting radioactive material, as is our focus.

As a result, research has grown in recent years to understand and detect CAN-based threats. Most network-based attacks can be classified into three categories:

- Fabrication: Injecting malicious CAN frames into the bus.
- Suspension: Silencing an ECU and preventing the transmission of frames on the bus.
- Masquerade: Advanced attacks include suspending a legitimate ECU's communication and replacing its message with a malicious payload.

This taxonomy is not all-encompassing and dismisses some types of attacks which may use non-standard attack vectors, e.g., application-layer exploits—such as address claim attacks seen in heavy-duty vehicles using SAE-J1939 CAN architecture [7].

Many researchers [8][9] have explored aftermarket plug-in devices to act as security appliances to detect CAN-based cyberattacks and improve vehicle cyber resilience. These devices are typically prototypes built with commercially available off-the-shelf single-board computers or microcontrollers—such as Raspberry Pi, Arduino, and Nvidia Jetson.

In the following section, we will describe a general development pipeline for in-vehicle CAN-based security hardware—namely, algorithm design, software and electronics development, hardware-in-the-loop integration, and system deployment—and identify their respective challenges. Furthermore, as our main contribution, we include our solutions, identified through practical experience, that expedite and facilitate this research to prototype process.

**IN-VEHICLE CYBERSECURITY RESEARCH, DEVELOPMENT, AND CHALLENGES**

Algorithm Research and Development
Algorithm development occurs as a requirement to detect new or unique classes of attacks. Intrusion detection systems (IDSs) are an effective and practical way to defend against an adversary and identify if a network is being attacked. Open-source datasets are particularly useful at this stage of development to quickly assess the behavior of the detection algorithm [10][11]. Fabrication and suspension attacks can be accurately detected in near real-time using frequency-based approaches (e.g., by simply modeling the AID timing characteristic). CAN messages are transmitted at a fixed frequency; thus, maliciously injected frames can be detected by enumerating messages within a time window. Python packages for data analysis are used to design and validate this approach, including NumPy, pandas, and scikit-learn. For instance, it is essential to compare competing methods to determine the best approach to a detection algorithm. In the case of fabrication attacks, we compared methods that were amenable to deploying the algorithm in real-time networks. This began as a proof of concept with a single attack on one vehicle [12]. Once the approach was validated, consideration was made to determine the most effective implementation for modeling the frequency of messages. In this case, we compared four competing methods to identify the best approach for in-vehicle deployment in operation environments [8].

Due to their complexity, masquerade attacks require a different approach. Since this class of attack does not impact the frequency of messages, it is a desirable candidate to explore machine learning-based detection methods. One approach is to analyze signals within CAN messages as they change over time. This type of signal-based intrusion detection is attractive due to the availability of machine-learning techniques for time-series data. For instance, Hanselmann et al. propose an ML-based method to detect masquerade attacks using deep-learning [13]. Their approach was tested on real and synthetic attack datasets using signals from a handful of AIDs. Moriano et al. [14] prescribes a solution using hierarchical clustering to detect similarities between CAN signals that can be potentially scaled up to hundreds of AIDs. Once an algorithmic approach is identified and tested, it must be integrated into a proper software engineering environment that provides control, infrastructure, and alerting mechanisms. Further sections will describe the integration of algorithms into software and hardware.

Software Engineering
While the *algorithm research and development* process creates the initial approach to detect a CAN-based cyberattack, software engineering makes the algorithm deployable and usable. Required components to deploy a detection algorithm in a realistic environment include responsive, alert handling, accurate and reliable logfile generation, and toolchains to support software updates and ensure programming code resilience.

Our software engineering process has three phases: *design, testing, and revision.* In the *design* phase, the infrastructure and supporting modules are created to support the deployment and maintenance of the detection algorithm. Three critical modules developed in this stage are logging, alerting, and error handling. The algorithm needs to generate an alert when an attack is detected appropriately, log relevant data concerning the attack, and safely handle software and hardware failures. The second phase, *testing*, ensures the software meets the needs and requirements to detect the attack properly. This process should ensure that the software does not contain errors and that the detection software works appropriately by evaluating the performance against pre-recorded network logs in software-simulated CANs. The detector is programmed to operate using real-time streaming data in this stage of development. It is vital to consider unusual vehicle environments for these types of tests—such as traffic recorded during maintenance or periods of inactivity. It is also essential to include a variety of attack examples to ensure the algorithm behaves as expected in its initial programming. Finally, the *revision* phase analyzes the testing results and addresses uncovered problems and priorities.

Note that these phases are not a simple sequence. Instead, they are a continuously iterative process that allows software engineers to develop functional software in the real world of changing requirements, hardware, and technical hurdles.

Hardware Engineering
The hardware used for rapidly prototyping in-vehicle security technologies includes embedded computers or microcontrollers and their accompanying electronics. The ubiquity of commercially available embedded single-board computers such as Raspberry Pi and Nvidia Jetson has dramatically accelerated the development of third-party embedded security tools by exposing general-purpose input/output capabilities with high-level programming language support, such as Python. These devices include a complete operating system, typically Linux-based, with a robust open-source and accessory market that provides many sensing and communication modules. Further, embedded computers such as the Nvidia Jetson line of products offer AI-accelerated processing to enable the deployment of machine learning models in embedded systems.

By contrast, microcontrollers are low-priced, simple devices designed for embedded tasks. Microcontrollers do not include an operating system and require a host computer for initial programming and debugging. Their computational abilities and power requirements are typically smaller than embedded computers. Consumer microcontrollers can vary in price and performance, but many popular architectures, such as Arduino, enjoy an active and mature open-source development community for software and hardware.

**Table 3. Common embedded hardware devices***

| Device Type | Device Name | Subcategory | Computational Performance | Power Requirements | Cost |
|---|---|---|---|---|---|
| **Embedded Computers** | Nvidia Jetson AGX | AI-accelerated embedded computer | Up to 12-core Arm Cortex-A78AE v8.2 64-bit CPU, 32GB LPDDR5 RAM | 10W-30W | $1000-$2000 |
| | Raspberry Pi 4 | Single-board computer | BCM2711, Quad core ARM Cortex-A72, up to 8GB LPDDR4 SDRAM | 15W | $35-$75 |
| | Raspberry Pi Zero 2 | Single board computer | 1GHz quad-core ARM Cortex-A53 CPU, 512MB SDRAM | 1.7W | $15-$30 |
| **Microcontrollers** | Teensy4.0 | Arduino-based microcontroller | ARM Cortex-M7 at 600 MHz, 1024K RAM | 0.5W | $15-$30 |
| | Arduino Due | Arduino-based microcontroller | 32-bit ARM at 84Mhz, 96 Kbytes of SRAM | 1W | $15-$30 |

* Manufacturer Suggested Retail Price as of May 2023.

When designing the hardware platform, three critical characteristics for deployment must also be considered—computational requirements, power consumption requirements, and dollar cost. Microcontrollers are often inexpensive and have low power consumption requirements but cannot run computationally expensive intrusion detection approaches. Embedded computers come at a slightly higher financial cost and computational ability but require more power. Finally, AI-accelerated embedded computers have tremendous processing ability but are extremely expensive and require the most considerable power to use their fullest processing power. It is important to consider power consumption requirements for the chosen hardware, as it can completely discharge automotive batteries if the vehicle is parked for a long time.

Hardware-in-the-loop Integration
This stage in the development pipeline focuses on preparing the detection appliance for in-vehicle deployment. One challenge when developing CAN-based security tools is the need for real data transmitted using CAN hardware. Obtaining an actual vehicle, while accurate, can be cost-prohibited and unsafe for repeated cybersecurity tests. As a result, constructing a testbed for hardware-in-the-loop integration is valuable to alleviate these common challenges.

We created a testbed that is reconfigurable, accessible remotely and retains physical-layer properties of CAN transmission. This system includes numerous Raspberry Pi 4 single-board computers with CAN add-on hats [15] representing simulated ECUs. These simulated ECUs communicate using the open-source Linux utility, *SocketCAN*. In addition to virtual CAN interfaces, SocketCAN includes a set of open-source user utilities named *"can-utils"* that provide a standard interface for interacting with CANs. Can-utils provides the following valuable tools: *candump, cansend, cangen,* and *canplayer*; Table 3 below summarizes these critical tools and their usefulness. To create a realistic environment with numerous ECUs transmitting simultaneously, messages within the pre-recorded log file are distributed across the simulated ECUs based on AID. Simulated ECUs use the canplayer tool to replay CAN traffic logs generated from experimentation.

**Table 3.** *Can-utils key tools*

| Tool | Description |
|---|---|
| *candump* | Displays and saves received CAN messages in a human-readable format. |
| *cansend* | Sends a single CAN frame to the specified CAN interface. |
| *cangen* | Generates random traffic on a specified CAN interface, useful for testing. |
| *canplayer* | Replays CAN log files on a CAN interface, simulating traffic. |

The testbed is managed by a central gateway allowing remote access and management of the simulated ECUs and prototype intrusion detectors. The prototype hardware is connected via CAN to the simulated ECUs, which can replay data pertinent to the type of testing required. An example of our testbed is depicted in Figure 4 below. This testbed is housed in a server rack, with the access gateway featured at the top. Three Raspberry Pi 4 single-board computers are inside the sliding drawer, acting as simulated ECUs. These devices are connected via CAN to two Nvidia Jetson Xavier AGX embedded computers, the prototype intrusion detection devices.

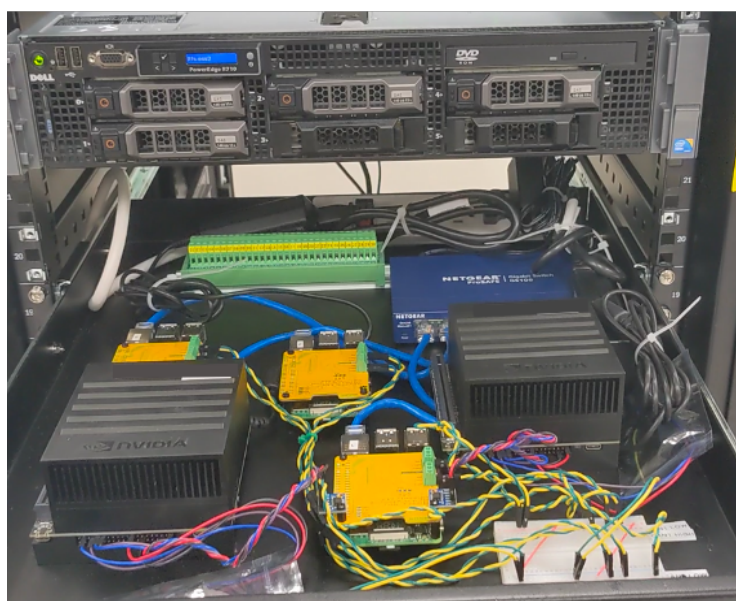

**Figure 3. CAN-based hardware integration testbed depicted.**

Once software and hardware perform sufficiently with no errors, the prototype detectors must prepare for in-vehicle deployment. In the following section, we will discuss challenges and considerations for the final phase of the development pipeline.

Prototype Deployment

Deploying the intrusion detection device in a limited pilot will ensure that the software, hardware, and algorithms behave appropriately when this tool is used in a production environment. The chosen vehicle or vehicle types for deployment will have unique challenges concerning power access, CAN access, and physical space. Automobiles can also be exposed to environments that harm electronics—such as extreme heat, humidity, or vibration. We can summarize these needs as functional (required for the operation of the device) or resilient (required for the device's health).

Functional requirements for prototype deployment enable the hardware and software to operate as designed and intended in-vehicle. This includes:

1. Proper power delivery: Most automobiles include a 12v battery accessible from the onboard diagnostic (OBD) connector. It is essential to understand the power requirements of the hardware platform, as some battery conditioning will be required to step voltages down or up from this source to provide proper power to the prototype detector. As noted in the *hardware engineering* subsection, since this battery supply is always active, it can completely discharge an automotive battery by not correctly preparing and managing the power draw. This may be catastrophic if the vehicle is required to move.
2. Adequate access to CAN(s): Modern automobiles may have numerous CANs within the vehicle. Networks are commonly exposed at the OBD connector, but it may be desirable to monitor and protect internal networks. Further, CAN bitrates can vary per vehicle and implementation. Automatic CAN bitrate detection is possible by connecting to the network at common bitrates (250kbps, 500kbps, 1000kbps) and observing the network traffic. The next bitrate value should be attempted if error frames or garbled messages are observed. It is important to note that the prototype device must not be programmed to broadcast CAN frames when attempting automatic bitrate detection. If an unsilenced device connects to a CAN with an improper bitrate, it will flood the network with errors and break communication between all nodes on the bus.
3. Safe and reliable in-cab mounting: The mounting location of the prototype will vary depending on the hardware construction and type of vehicle chosen for deployment. Smaller microcontrollers can be more versatile in their mounting and placement options. This type of device is a popular choice near the OBD plug or secured onto internal body panels. For larger constructions, there are more restrictions for placement. Good locations may include trunks, seat undersides, and storage bins. Notably, this will remove functional space and may inconvenience drivers or passengers.

Resilient requirements for prototype deployment include those which ensure the health and longevity of the device. These are primarily environmental considerations and include the following:

1. Temperature: Extreme temperatures and rapid fluctuations impact the performance and longevity of electronics. Previous research has observed that the in-cabin temperatures for a stopped vehicle can reach more than 140 degrees Fahrenheit within 20 minutes of stopping [16]. If the climate control cannot cool inside the truck cabin, the processor and onboard components will overheat, potentially damaging the device. Ensuring the prototype is outside direct sunlight and considering active cooling solutions is critical. Still, it may be impossible to cool the device for safe operation during extremely high temperatures; thus, it is crucial to establish overheating shutdown protections or failure modes.
2. Vibration and movement: Radiological transportation occurs in various environments, including roads that may not be wholly maintained due to weather or other conditions. Further,

   some routes may require periods of driving on unpaved or gravel roads. Thus, it is essential to consider secure mounting and cable management of the prototype detector. It could be dangerous to the driver or vehicle if the device jostles loose or cables snag internal components.
3. Security: For resilience, it is imperative to consider the security implications the prototype detector may bring. If an embedded computer is deployed, it is advised to disable wireless networking unless required and to maintain updated operating system security patches. For initial prototype deployments, it is suggested to physically alter the connection between the CAN controller and CAN transceiver to remove the node's ability to transmit messages, as this eliminates the use of the prototype device from (inadvertently or being leveraged adversarially in) attacking the CAN.

Real-world experiments can verify the approach once the intrusion detection tool is adequately powered, connected to the automotive network, and confidently shielded from the environment. This deployment aims to understand and remedy issues uncovered during initial piloting. Further refinement then occurs, which can include the maturation of this technology into a field-deployed intrusion detection system with full-time monitoring capabilities.

**CONCLUSIONS**
Developing and deploying in-vehicle security applications is a nontrivial pursuit with continual challenges as new electronics are introduced into automobiles. This publication has detailed Oak Ridge National Laboratory's approach to rapidly prototype these tools. Our process begins with understanding the data and a desire to detect new, novel attacks and threats. The proliferation of machine-learning technologies enables the creation of advanced detection algorithms for multiple classes of attack vectors—known and unknown. Algorithms are not useful on their own; thus, hardware and software infrastructure is created so the algorithm can detect attacks, which generate alerts, and provide metadata and logs relevant to the attack period. Proper hardware selection must consider the computational limits of differing platforms, the targeted vehicle type, and constraints when deploying the hardware in-vehicle. Finally, all the components are thoroughly tested and integrated through a hardware-in-the-loop testbed which includes simulated ECUs and remote access. Once the prototype intrusion detector performs sufficiently with no errors, the deployment phase begins—the long-term mounting and testing of the device on actual vehicles in real-world, unpredictable environments. Further refinement continues indefinitely until the project reaches its conclusion. Overall, this process has dramatically accelerated the development time from proof of concept to deployed in-vehicle prototype and allows for the concurrent completion of milestones.

**ACKNOWLEDGMENTS**

The authors sincerely thank Bobby Bridges, Mingyan Li, and Shannon Morgan for contributing to transportation cybersecurity research.